\documentclass[letterpaper, 10 pt, conference]{ieeeconf}  

\IEEEoverridecommandlockouts                              

\usepackage{graphics} 
\usepackage{epsfig} 
\usepackage{mathptmx} 
\usepackage{times} 
\usepackage{amsmath} 
\usepackage{amssymb}  
\usepackage{circuitikz}
\usepackage{hyperref}
\let\labelindent \relax

\usepackage{enumitem}

\usepackage{tikz-cd}
\usepackage{subcaption}
\usepackage[nameinlink, capitalise]{cleveref}
\usetikzlibrary{shapes.misc}
\usepackage{float}

\title{\LARGE \bf Towards Statistical Methods for Minimizing Effects of Failure Cascades}

\author{Siyu Liu$^{1}$ and Marija Ilic$^{2}$
\thanks{$^{1}$Massachusetts Institute of Technology, 77 Massachusetts Ave, Cambridge, MA 02139, USA
{\tt\small eliu24@mit.edu}}%
\thanks{$^{2}$MIT Laboratory for Information and Decision Systems, 77 Massachusetts Ave, Cambridge, MA 02139, USA
        {\tt\small ilic@mit.edu}}%
}

\begin{document}
\maketitle
\thispagestyle{empty}
\pagestyle{empty}

\begin{abstract}
This paper concerns the potential of corrective actions, such as generation and load dispatch on minimizing the effects of transmission line failures in electric power systems. Three loss functions (grid-centric, consumer-centric, and influence localization) are used to statistically evaluate the criticality of initial contingent failures. A learning scheme for both AC and DC grid models combine a Monte Carlo approach with a convex dynamic programming formulation and introduces an adaptive selection process, illustrated on the IEEE-30 bus system.
\end{abstract}

\section{INTRODUCTION}

Modern power systems are prone to many unpredictable component failures. Contingencies caused by failed transmission lines, if not treated promptly, can propagate to other system components and cause large scale blackouts \cite{bg-1, bg-2, bg-3}, incurring immeasurable economic loss \cite{yamashita, puerto_rico}.

Past events have shown that large scale blackouts are typically results of sequential failures of transmission lines, called \textit{failure cascades.} Analyzing these cascades is computationally expensive, given the scale and complexity of electrical networks that can have hundred of thousands of components with nonlinear dynamic interactions and time-varying parameters. Additionally, cascades tend to evolve quickly, leaving only up to 15 minutes for the system operators to take corrective actions before the failure propagates \cite{puerto_rico}. As such, it is particularly important to understand the cascade patterns and their induced effects, in order to advice system operators during such extreme events. 

Most of the early work concerns analysis of cascading failures \cite{earlyworks}. In this paper, the main problem of interest is how to go beyond analysis and introduce statistical tools for minimizing the effects of different equipment failures on grid integrity and the ability to serve customers as the events take place. Statistical knowledge is utilized to quickly identify the most likely effects and compute the corrective actions (which we use to refer to generation re-dispatch and preemptive load shed) needed to prevent future spreading of the failures while serving the grid as much as possible. 

In previous studies, common approaches to predict the cascade sequence all require solving the power flow problem. This is especially difficult to implement with the AC power flow (AC PF) model due to its high computation cost and convergence problems \cite{allen, vaiman, history}. As such, most offline statistical analysis has been done with DC power flow (DC PF) \cite{Wu}. However, DC models under-estimate the effects of failures and do not always provide physically implementable solutions because they do not consider reactive power and voltage constraints \cite{Cetinay}. To overcome the high complexity of flow analysis, researchers sought various statistical methods to build simpler approximations, including random graphs, branching processes, and flow dynamics models --- all typically embed non-trivial constraint relaxations \cite{song}-\cite{bg-zhang2}. 

To remedy these shortcomings, we adopt a flow-free approach using the influence model (IM). The IM can be seen as a Markovian model that, at each time step, gives the probability of each link failing given the current network profile. It is straightforward to construct and fast to implement for large-scale applications. It reduces the task of line failure and mandatory load shed prediction to matrix multiplication, completely eliminating the burden of flow computation in real time. We borrow insights from previous works to train the model using Monte Carlo, convex optimization, and adaptive selection, while drastically augment the scope in both methodology and results \cite{Wu, song}.

However, most previous studies on link failures only analyze the effects of initial contingencies and do not consider the effects of corrective actions. Those that do are done with the DC model, which do not render physically implementable \cite{correction1, correction2}. In terms of load shed prediction, different methods have been proposed to find the optimal load shedding schedule subject to load priority, but only for the purpose of minimizing grid-centric cost \cite{sinha, r-n, ShiLiu}. Others including \cite{Cetinay} and \cite{Wu} implement a constant factor load shed algorithm without any ``smart scheduling.'' To the best of our knowledge, there is so far no statistical model for load shed prediction. 

In this paper, \cref{section 2: matrix D} and \cref{section 3: matrix E} introduces the IM for link failure and load shed prediction. \cref{section 4: sample pool} outlines our simulation process. \cref{section 5: corrective actions DC} proposes three evaluation metrics and use them to evaluate corrective actions in DC models, and \cref{section 6: corrective actions AC} repeats this analysis in AC models. \cref{section 7: performance eval} summarizes the prediction accuracy and time complexity of our models. \cref{section 8: advisory tool} underlines the practicality of our work by applying it in industry.

\section{THE INFLUENCE MODEL $(A^{11}, A^{01}, D, \epsilon)$ FOR CASCADE PREDICTION} \label{section 2: matrix D}
The influence model is a Markovian-like model whose dynamics are described by the state variable transitions. The model operates on a network state vector, $s,$ to predict subsequent network states. $s$ is a $(N_{br}\times 1)$ vector that stores the status of all links in binary, where $N_{br}$ is the number of transmission lines in the network. Given the $i$-th link, $s_i[t]=1$ indicates that link $i$ is alive at time $t,$ and $s_i[t]=0$ indicates that it has failed. We define $s:=\{s[t]\}_{t=0}^T$ as the state sequence with $T$ being the termination time of the cascade, where either the cascade has stopped propagating or all links have failed.

The influence model is defined by transition probability matrices $A^{01}$ and $A^{11},$ and the weighted influence matrix $D.$ Particularly, the $ji$-entry of $A^{11}$ and $A^{01}$ denotes the probability of link $i$ being alive given that link $j$ is, respectively, dead or alive in the previous time step. Explicitly, 
\begin{equation} \label{eq:A11}
    A^{11}_{ji}:=\mathbb{P}(s_i[t+1]=1|s_j[t]=1),
\end{equation}
\begin{equation} \label{eq:A01}
    A^{01}_{ji}:=\mathbb{P}(s_i[t+1]=1|s_j[t]=0).
\end{equation}

Combining the transition probabilities from $A^{11}$ and $A^{01}$ and weighing them by the influence factor provided by matrix $D,$ the transition of a state variable $s_i$ is given by
\begin{equation} \label{eq:s[t+1]}
    \widetilde{s_i}[t+1] = \sum_{j=1}^{N_{br}} d_{ji}\left(A^{11}_{ji}s_j[t]+A^{01}_{ji}(1-s_j[t])\right).
\end{equation}

The weight $d_{ij}$ represents the proportional affect from the link $j$ to $i,$ under the constraint that $\sum_{j=1}^{N_{br}} d_{ij} = 1$ and $d_{ij}\geq 0$ for all links $i, j.$ It can be interpreted as the ratio of the influence from link $j$ to $i$ among all links over $i.$ $\widetilde{s_i}[t+1]$ is the predicted probability of the link $i$ being normal at time $t+1$ given the network state $s[t].$ Specifically, suppose link $j$ is normal at time $t,$ the term $(A^{11}_{ji}s_j[t]+A^{01}_{ji}(1-s_j[t])$ in \cref{eq:s[t+1]} becomes $A^{11}_{ji},$ which is the conditional probability of link $i$ being normal at $t+1$ given $j$ is normal at $t.$ This method is certified physically meaningful in a 3-bus system.

After $\widetilde{s_i}[t+1]$ is obtained, we apply a deterministic bisection scheme to make a prediction for the state of link $i$ based on a threshold $\epsilon_i$. In particular, we predict the state of link $i$ to be healthy if $\widetilde{s_i}[t+1]\geq \epsilon_i,$ and out of service otherwise. 

The next three subsections are dedicated to explaining how we learn the values of $A^{11}, A^{01}, D,$ and $\epsilon.$ 

\subsection{Matrices $A^{11}$ and $A^{01}$ for Link-on-Link Influence} \label{section 2.2 estimating A}

We obtain $A^{11}$ and $A^{01}$ with a Monte Carlo method. Let $\tau^k_i$ be the time step that link $i$ changes to failure state in the $k$-th cascade sequence $s^k.$ If link $i$ does not fail in $s^k,$ we set $\tau^k_i$ to be the termination time of $s^k.$ Then, we compute for the value of $A^{11}_{ji}$ by tallying the number of time steps that $j$ spent in the two states before link $i$ failed. In particular, \begin{equation}
    A^{11}_{ji} := \dfrac{\sum_{k=1}^K C^{11}_{ji}(s^k, \tau_i^k)}{\sum_{k=1}^K C^{1}_{j}(s^k, \tau_i^k)},
\end{equation}
where $C^{1}_{j}(s^k, \tau_i^k)$ is the number of time steps before $\tau^k_i$ in the $k$-th sample such that link $j$ is normal; $C^{11}_{ji}(s^k, \tau_i^k)$ is the number of time steps before $\tau_i^k$ in the $k$-th sample such that link $i$ is normal, given link $j$ is normal on the adjacent upstream time step.

Similarly, we compute $A^{01}_{ji}$ via
\begin{equation}
    A^{01}_{ji} := \dfrac{\sum_{k=1}^K C^{01}_{ji}(s^k, \tau_i^k)}{\sum_{k=1}^K C^{0}_{j}(s^k, \tau_i^k)},
\end{equation}
where $C^{0}_{j}(s^k, \tau_i^k)$ is the number of time steps before $\tau^k_i$ such that link $j$ has failed; $C^{01}_{ji}(s^k, \tau_i^k)$ is the number of time steps before $\tau_i^k$ such that link $i$ is normal, given link $j$ is failed on the adjacent upstream time step. Readers may refer to \cite{Wu} for a toy example. In our experiments, Monte Carlo counting starts at the first uncontrolled step.

\subsection{Weighted Influence Matrix $D$} \label{section 2.3 estimating D}
With $A^{11}$ and $A^{01},$ their values can be substituted into \cref{eq:s[t+1]} to form an optimization problem whose decision variables are only from $D.$ We choose the objective function to be the least square error function to formulate a convex quadratic optimization problem
\begin{equation} \label{eq: optimization D}
\min_D \frac{1}{K} \sum_{k=1}^K\sum_{t=1}^{T_k}\sum_{i=1}^{{N_{br}}} \left(s_i^k[t+1]-\sum_{j=1}^{N_{br}} d_{ji}\left(A^{11}_{ji}s_j[t]+A^{01}_{ji}(1-s_j[t]\right)\right)^2
\end{equation}
s.t. $\sum_{j=1}^{N_{br}} d_{ij} = 1 \forall i$ and $d_{ij}\geq 0 \forall i, j,$ where $k$ is the sample index, $t$ is the time step, and $i, j$ the bus indices. 

\subsection{Bisection Thresholds $\epsilon$} \label{section 2.4 epsilon}
In order to improve prediction accuracy, we determine the threshold value $\epsilon_i$ for each link under each contingency profile. Wu et al. have shown that naively choosing 0.5 as the threshold could introduce large prediction errors \cite{Wu}. The threshold $\epsilon_i$ is determined in three steps as follows.

Step 1: We identify the threshold value of link $i$ in each of the $K$ samples. Three scenarios can happen.
\begin{enumerate}[leftmargin=*]
    \item Link $i$ fails initially. In this case, we choose the threshold to be $1$ to denote default failure.
    \item Link $i$ fails at the $(t+1)$-th step (i.e. $s^k_i[t]=1,$ $s^k_i[t+1]=0$). We set the threshold value to be $\frac{\widetilde{s^k_i}[t]+ \widetilde{s^k_i}[t+1]}{2}.$
    \item Link $i$ never fails. In this case, we set the threshold value to be $\alpha_D \cdot \widetilde{s^k_i}[T_k],$ where $T_k$ is the lifetime and $\alpha_D$ a constant between 0 and 1.
\end{enumerate}

Step 2: We form the threshold pool, $\Omega^D_i,$ for each link $i.$ This threshold pool contains every threshold found in Step 1 and their corresponding initial contingency profiles, which include their loading levels and initial failures.

Step 3: We select the appropriate threshold value for link $i$ from $\Omega^D_i$ for a new contingency. To test a new contingency, we first determine the initial loading level, and eliminate all thresholds that does not have the same initial loading level. Then, we select the threshold value such that its corresponding initial failures is the ``closest'' to the new contingency $s^{new}[1].$ ``Closeness'' is measured by the L1 norm of $s^{new}[1]-s^k[1],$ which we select $k^*$ to minimize:
\begin{equation} \label{eq:k*}
k^* = \underset{k=1, 2, \dots, K}{\arg\max} ||s^{new}[1]-s^k[1]||_1,
\end{equation}
where $k$ is the index of the known contingency under the same loading profile. We select $\epsilon_i^{k^*}$ to be our threshold value. In the case that multiple solutions for $k^*$ exist for \cref{eq:k*}, we choose $\epsilon_i$ to be the median value among all $\epsilon_i^{k^*}$'s.

\subsection{Run Time Analysis} \label{section 2.5 D runtime}
We analyze the time complexity to train the influence model. To obtain influence matrices $A^{11}$ and $A^{01},$ the Monte Carlo process has a time complexity of $O({N_{br}}^2)$ for any number of samples. For each of the ${N_{br}}^2$ elements in each matrix, it takes $O(1)$ to determine $C^{01}_{ji}, C^{11}_{ji}, C^{01}_{j}, C^{1}_{j}.$ This run time was reduced from \cite{Wu} by a factor of $T$ (the cascade length). The optimization and selection steps to obtain $D$ and $\epsilon$ have complexities $O(M^4)$ and $O(MT),$ on par with \cite{Wu}.

\section{THE INFLUENCE MODEL $(B^{11}, B^{01}, E, \delta)$ FOR MANDATORY LOAD REDUCTION PREDICTION} \label{section 3: matrix E}
\subsection{Overview}
We propose an influence model for mandatory load shed prediction inspired by the link fail model in \cref{section 2: matrix D}. 

At any time $t,$ this model predicts the load binary vector $l[t]$ based on the network state $s[t].$ For each bus $i,$ the binary variable $l_i[t]$ denotes whether bus $i$ can be served in full as the state of the network goes from $s[t]$ to $s[t+1].$ $l_i[t]=1$ indicates full service, and $l_i[t]=0$ indicates load reduction. The collection of these variables defines the state sequence $l:=\{l[t]\}_{t=0}^{T-1},$ where $T$ is the cascade termination time.

This model consists of influence matrices $B^{11}, B^{01}, E,$ and the threshold vector $\delta.$ In particular, matrix $E$ of size $(N\times N_{br})$ defines the weighted influences from links to buses, where $N$ is the number of buses and $N_{br}$ the number of links. Each entry $e_{ij}$ denotes the proportional influence of link $j$ on bus $i,$ subject to $\sum_{j=1}^{N_{br}}=1$ and $e_{ij}\geq 0 \forall i, j.$ The influence matrices $B^{11}$ and $B^{01}$ of size $(N_{br}\times N)$ define the conditional probabilities of service reduction. Explicitly,
\begin{equation} \label{eq:B11_prob}
    B^{11}_{ji}:=\mathbb{P}(l_i[t]=1|s_j[t]=1),
\end{equation}
\begin{equation} \label{eq:B01_prob}
    B^{01}_{ji}:=\mathbb{P}(l_i[t]=1|s_j[t]=0).
\end{equation}

Combining the transition probabilities in $B^{11}$ and $B^{01}$ and weighing them by the influence factors from matrix $E,$ we can estimate the load state variable $l_i[t]$ by
\begin{equation} \label{eq: l[t]}
    \widetilde{l_i}[t] = \sum_{j=1}^{N_{br}}e_{ij}\left(B^{11}_{ji}s_j[t]+ B^{01}_{ji}(1-s_j[t])\right),
\end{equation}
where $j$ is the branch index and $s_j[t]$ the status of branch $j$ at time $t.$ The probability of the system being able to serve full load at bus $i$ is calculated by a weighted sum of the influence from all links. Note that in this calculation, we use the actual cascade sequence $s$ instead of our prediction $\widetilde{s}$. This is done so that the performance of the load shed prediction is independent from link failure prediction. So we can optimize for the two influence models concurrently. 

In the next step, we apply a bisection scheme to predict whether load shed occurred. Specifically, we predict that full load is served if $\widetilde{l_i}[t]\geq \delta_i,$ and load shed occurred otherwise.

The next three subsections provide a detailed explanation to how the influence model is obtained. This model is verified to be practically meaningful in a 3-bus system. 

\subsection{Matrices $B^{11}$ and $B^{01}$ for Link-on-Bus Influence}\label{section 3.2 B matricies}

Similar to \cref{section 2.2 estimating A}, we adopt a Monte Carlo approach to determine the link-on-bus influences. Entries $B^{11}_{ji}$ are defined by
\begin{equation}\label{eq:B11}
    B^{11}_{ji}:=\dfrac{\sum_{k=1}^K F^{11}_{ji}(k)}{\sum_{k=1}^K F^1_j(k)},
\end{equation}
where $F^{11}_{ji}(k)$ denotes the number of time steps where bus $i$ did not shed load given that link $j$ is alive in the adjacent upstream step, $F^1_j(k)$ denotes the total duration that link $j$ is normal in sample $k.$

We similarly define entries in $B^{01}:$
\begin{equation}\label{eq:B01}
    B^{01}_{ji}:=\dfrac{\sum_{k=1}^K F^{01}_{ji}(k)}{\sum_{k=1}^K F^0_j(k)},
\end{equation}
where $F^{01}_{ji}(k)$ denotes the number of time steps where bus $i$ did not shed load given that link $j$ is dead in the adjacent upstream step, $F^1_j(k)$ denotes the total duration that link $j$ is dead in sample $k.$

The key difference between the $A$ matrices and $B$ matrices is that, in the $A$ matrices, once a link has failed, it cannot become alive again, thus we do not need to consider the times steps after it terminating time, but for the $B$ matrices, each bus can shed load many times, regardless the evolving network state. Hence, we must consider all time steps up to network lifetime $T_k$ when estimating the $B$ matrices and optimizing for the $E$ matrix.

\subsection{Weighted Influence Matrix $E$} \label{section 3.3 E}

To obtain values in the weighted influence matrix $E,$ we use a similar approach as in \cref{section 2.3 estimating D} to optimize for the $(N\times N_{br})$ entries as decision variables, as formulated in
\begin{equation} \label{eq: optimization E}
\min_E \frac{1}{K} \sum_{k=1}^K\sum_{t=1}^{T_k}\sum_{i=1}^{N} \left(l_i^k[t]-\sum_{j=1}^{N_{br}}e_{ij}\left(B^{11}_{ji}s_j[t]+ B^{01}_{ji}(1-s_j[t])\right)\right)^2,
\end{equation}
s.t. $\sum_{j=1}^{N_{br}}e_{ij}=1$ and $e_{ij}\geq 0$ for all $i, j,$ where $k$ is the sample index, $t$ the time step, $i$ the bus index, and $j$ the link index.

\subsection{Bisection Thresholds $\delta$}\label{section 3.4 delta}
We similarly determine the bisection thresholds $\delta_i$ for each bus $i.$ This threshold is determined in the following 4 steps.

Step 1: We identify $\delta_i$ for each sample. For any bus $i$ in sample $k,$ four scenarios could arise. 
\begin{itemize}[leftmargin=*]
    \item Bus $i$ never sheds load. We let $\delta_i = \alpha_E\cdot \min_{t\in [1, T_k]},$ where $\min_{t\in [1, T_k]}$ selects for the minimum probability of full service and $\alpha_E$ is some constant between 0 and 1. 
    \item Bus $i$ sheds load at every step. We set $\delta_i = 1.$ In the case that all links fail immediately after the initial contingency (i.e. $T_k = 1$), we set $\delta_i = 0.5 \cdot (1+\widetilde{l_i^k}[1]).$
    \item In all other cases, where bus $i$ sheds load some times and serves full load at other times, we adopt a conservative approach by setting $\delta_i = \min(P, Q),$ where $P:=\max\{\widetilde{l_i}[t]|l_i[t]=0\}$ (the maximum predicted probability where there is no load shed) and $Q:=\min\{\widetilde{l_i}[t]|l_i[t]=1\}$  (the minimum predicted probability where there is load shed), as it is less costly to be over-cautious.
\end{itemize}

Step 2: We collect the contingency profile threshold for bus $i$ from every sample in the threshold pool $\Omega^E_i.$

Step 3: To predict load sheds under a new contingency, we first select the appropriate threshold from the pool $\Omega^E,$ following the exact same procedure as \cref{section 2.4 epsilon}, and determine their median if multiple thresholds are selected.

\subsection{Run Time Analysis} \label{section 3.5 E run time}
The run time for building our model is dominated by the optimization step to obtain $E.$ Evaluating the $B$ matrices takes $O(N N_{br})$ time. The optimization step takes $O(N^2 N_{br}^2),$ as the size of the decision variables set is $N\times N_{br},$ each taking $O(N\times N_{br})$ to optimize. Determining thresholds takes $O(NT),$ where $T$ is the maximum length of the cascade. 

The time complexity using this model for prediction will be analyzed in \cref{section 7: performance eval}.

\section{SAMPLE POOL GENERATION} \label{section 4: sample pool}

In this section, we discuss how to generate the sample pool. As there is no standard oracle for assessing failure cascades at present, we base our experiments on the Cascading Failure Simulator (CFS) proposed in \cite{CFS}, as did Wu et al.\cite{Wu}. The CFS oracle is similar to the short-term ORNL-PSerc-Alaska (OPA) oracle, except that line outages are treated deterministically and does not apply optimal re-dispatch during a failure cascade. Instead, it only sheds load or curtails generation if system-wide power mismatch occurs. For samples where no corrective actions are taken, we follow the CFS oracle exactly, and for sample where corrective actions are applied, we follow a relaxed version of the CFS oracle by allowing re-dispatch during the cascade. This relaxation is realistic, as the time between two failures can be as long as 15 minutes to allow the re-dispatch \cite{puerto_rico}. In all our experiments, we initialize the network as fully functional, randomly select two initial contingencies, and determine the cascade sequence following the oracle. Long term thermal condition is used when all links are fully functional, and it changes to short term thermal conditions once failures occur (which we assume to be $1.05\times$ long term). After each failure, we solve the DC/AC PF/OPF problem using the MATPOWER Toolbox \cite{MATPOWER}. Our three sets of experiments are defined with parameters as follows.

\textit{Experiment 1: No corrective action.}
In this experiment, we do not execute upon contingencies. We simply record the network status and loading levels at each bus at each step of the cascade.

\textit{Experiment 2: Generation re-dispatch for full service.}
We act upon contingencies by re-dispatching generation. We solve OPF under both uniform generation cost and bus-specific generation cost provided by \cite{MATPOWER}, and compare their results. We aim to serve all loads in full and only shed load uniformly in scale when unable to serve in full.

\textit{Experiment 3: Generation re-dispatch with smart service reduction.}
This experiment resembles \textit{Experiment 2}, except that instead of aiming for full service, we find the OPF solution that minimizes cost of shedding load. We assume cost of load shed at each bus is uniform or priority-based. Note that in this experiment, no links will fail, as the optimization is solved subject to link constraints, and the maximum service if part of the optimization objective.

\begin{figure}[hbtp]
\centering
\includegraphics[width=0.25\textwidth]{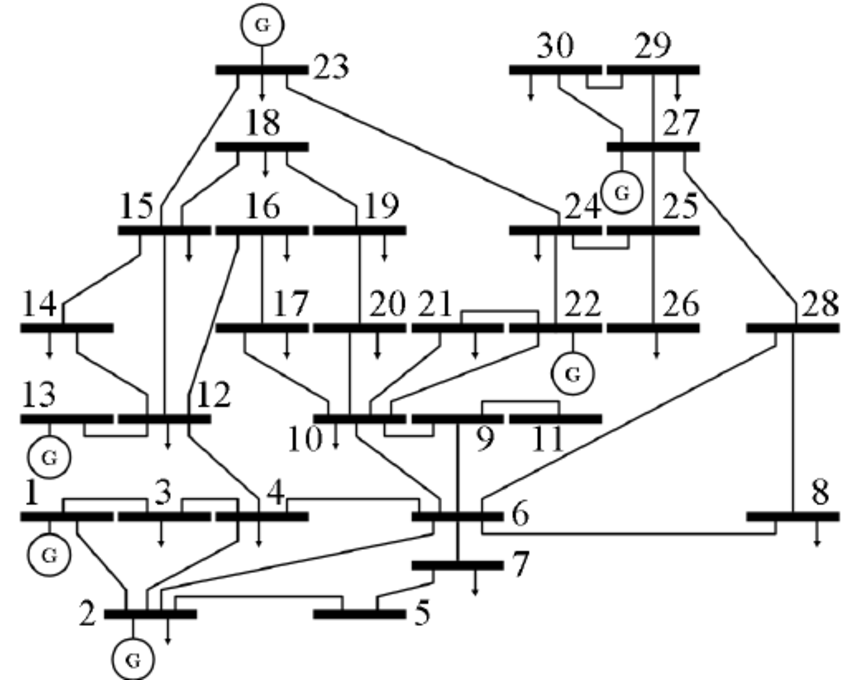}
\captionsetup{font={footnotesize}}\caption{Single-line diagram of IEEE 30-bus system. The bars represent the buses; the circles represent generators. The arrows represent loads. The lines connecting buses represent the transmission lines \cite{one-line-diagram}.} \label{fig: IEEE30 one line}
\end{figure}

Our experiments are executed on the IEEE-30 system with initial loading being $c$ times the test case from \cite{MATPOWER}. The $c$ ranges from $0.9$ to $1.8$ in $0.1$ increments. The upper bound $1.8$ is chosen because the maximum total generation is $1.77\times$ the default total load. \cref{fig: IEEE30 one line} provides a one-line diagram to visualize the IEEE-30 network.

\section{EFFECTS OF CORRECTIVE ACTIONS (DC MODELS)} \label{section 5: corrective actions DC}

We examine the effects of corrective actions under the DC model. We propose three loss functions as benchmarks to quantify grid-centric and consumer-centric loss, as well as influence localization in each failure cascade. Comparison between scenarios is done by taking the mean cost of all 300 Monte Carlo trials. Graphs in \cref{fig:all plots} show the loss functions evaluated at different loading levels for all three experiments (six set ups in total) over increasing loading levels.

Our experiment finds that, under certain parameters, full service is impossible even without contingencies, as our solver fails to converge. The problem arises under scenarios at high loading levels or non-uniform bus priorities. In \textit{Experiment 2}, DC OPF and AC OPF fail to converge for loading levels greater than $1.3 \times$ and $1.0\times$ the default loading level, respectively. This signifies the necessity for smart service reduction. The rest of this section presents highlights of the analysis of runs where initialization succeeds.

\subsection{Grid-Centric Loss} \label{section 5.1: link fail DC}
For each cascade sample, grid-centric loss is defined as
\begin{equation} \label{eq: link fail loss}
    Link Fail Loss = \sum_{b=1}^{N_{br}} C(b)\cdot e^{-0.2t_b},
\end{equation}
where $C(b)$ is defined as the cost on branch $b,$ proportional to its maximum thermal capacity, and $t_b$ is the life time of $b.$ The discounting factor $e^{-0.2t_b}$ is to penalize early failures.

In \textit{Experiment 1 \& 2}, links fail more frequently and earlier on in the cascade at higher initial loading levels. Even in the only two loading levels where \textit{Experiment 2} only successfully initializes, the loss of link failure is much greater than that in \textit{Experiment 1}. This demonstrates that PF models underestimates failure sizes. There is no observable difference between re-dispatching with actual or uniform generation cost in \textit{Experiment 2 \& 3}. \cref{fig: linkfail DC} illustrates these results.

\subsection{Consumer-Centric Loss} \label{section 5.2: load shed DC}
For each cascade sample, consumer-centric loss is defined with the formula as follows.

\begin{equation}\label{eq: load shed loss}
    Load Shed Loss = \sum_{l=1}^{N}\sum_{t=1}^{T_k-1} C(l) \cdot LS_l(t) e^{-0.2t},
\end{equation}
where $C(l)$ is denotes the load priority and $LS_l(t)$ the amount of load shed between time steps $t$ and $t+1$ at bus $l.$ The expression is similarly discounted by $e^{-0.2t}$ to signify the diminishing severity over time.

Our experiments yield one notable finding. If corrective actions are taken promptly, we may preserve infrastructure integrity in full without significant service reduction. In particular, load shed loss is minimized under \textit{Experiment 3}'s set up, when we run OPF with cost-based load shed. As such, the flow on all links are within their capacities and no cascade incurs. As illustrated in \cref{fig: loadshed DC}, comparing the load shed across all three experiments, \textit{Experiment 3}'s design reduces the consumer-centric loss to much less than that in \textit{Experiment 1 \& 2}. The passive, emergency load shed in \textit{Experiment 2} incurs the greatest loss. Whether the cost of generation varies at different buses does not induce significant difference in the load shed loss. 

The only challenge to implementing this solution is that this re-dispatch must finish between the time of the initial contingencies and any subsequent contingencies. Depending on the extent of congestion, electrical equipment can operate over capacity for 5 to 15 minutes \cite{puerto_rico}. As long as re-dispatching within this 5-15 minute window, this corrective strategy can fully halt the cascade while providing maximum service to grid users.

\subsection{Localizing Influence} \label{section 5.3: local influence DC}
We can evaluate the level of localization of influence under each topology and initial loading condition through the $D$ matrices. Here, we define \textit{localization} as \textit{only affecting links that are physically nearby.} As \cite{Wu} has shown, the network's electrical properties can create congestion far away from the failed links. Local influence loss is defined with the following equation.
\begin{equation} \label{eq: local influence loss}
    Local Influence Loss = \sum_{n_1=1}^{N_{br}}\sum_{n_2=1}^{N_{br}} D_{n_1, n_2}\cdot K(n_1, n_2).
\end{equation}
where $K(n_1, n_2)$ the distance between links $n_1$ and $n_2,$ defined as the geodesic distance between vertices $v_1, v_2$ that correspond to $n_1, n_2$ in the line graph of the network.

We expect contingencies easier to control in localized networks: if initial failures only affect lines that are nearby, these nearby failures can create islands to contain the failure.

\begin{figure}[hbtp]
\begin{minipage}{0.24\textwidth}
\includegraphics[width=1\textwidth]{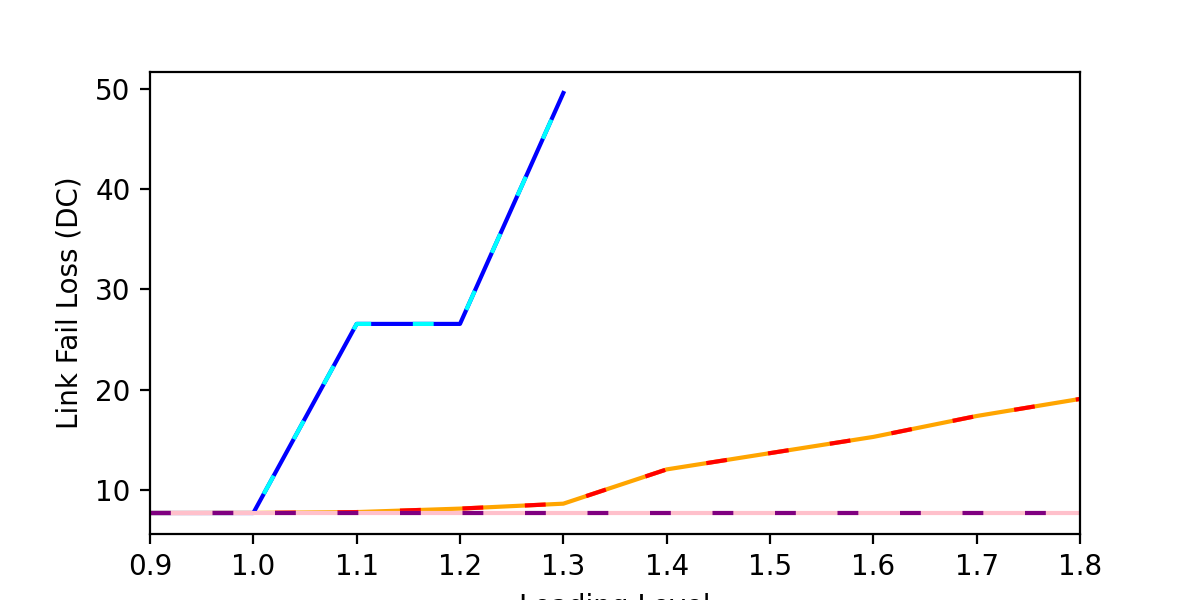}
\captionsetup{font={footnotesize}}\caption{Link Fail Loss (DC).} 
\label{fig: linkfail DC}
\end{minipage}\hfill
\begin{minipage}{0.22\textwidth}
\includegraphics[width=1\textwidth]{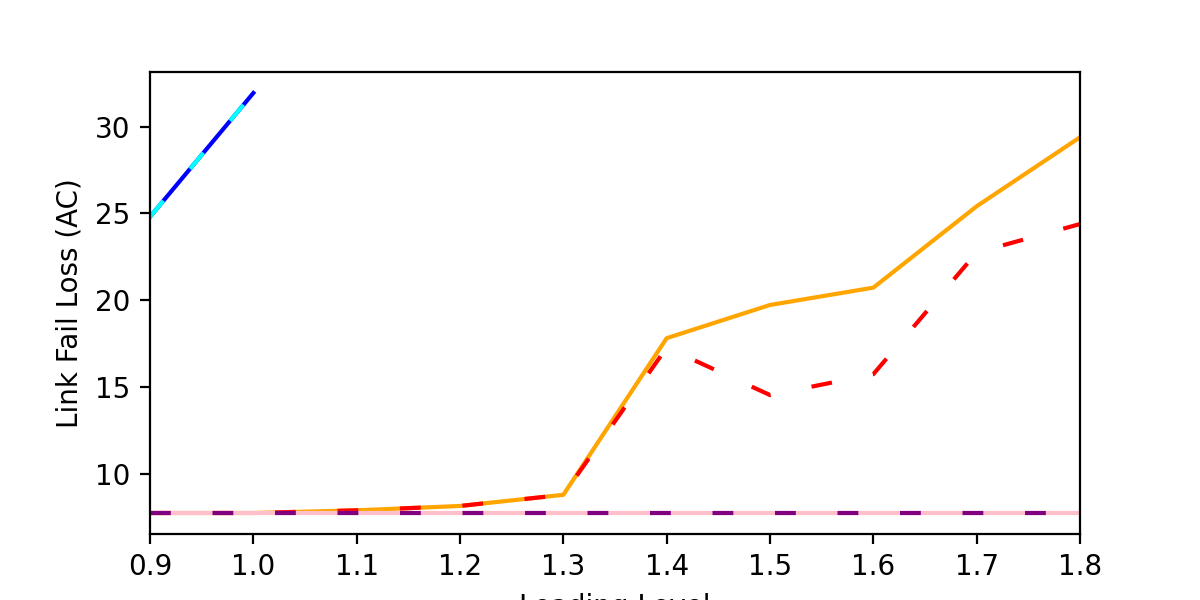}
\captionsetup{font={footnotesize}}\caption{Link Fail Loss (AC).}
\label{fig: linkfail AC}
\end{minipage}
\begin{minipage}{0.22\textwidth}
\includegraphics[width=1\textwidth]{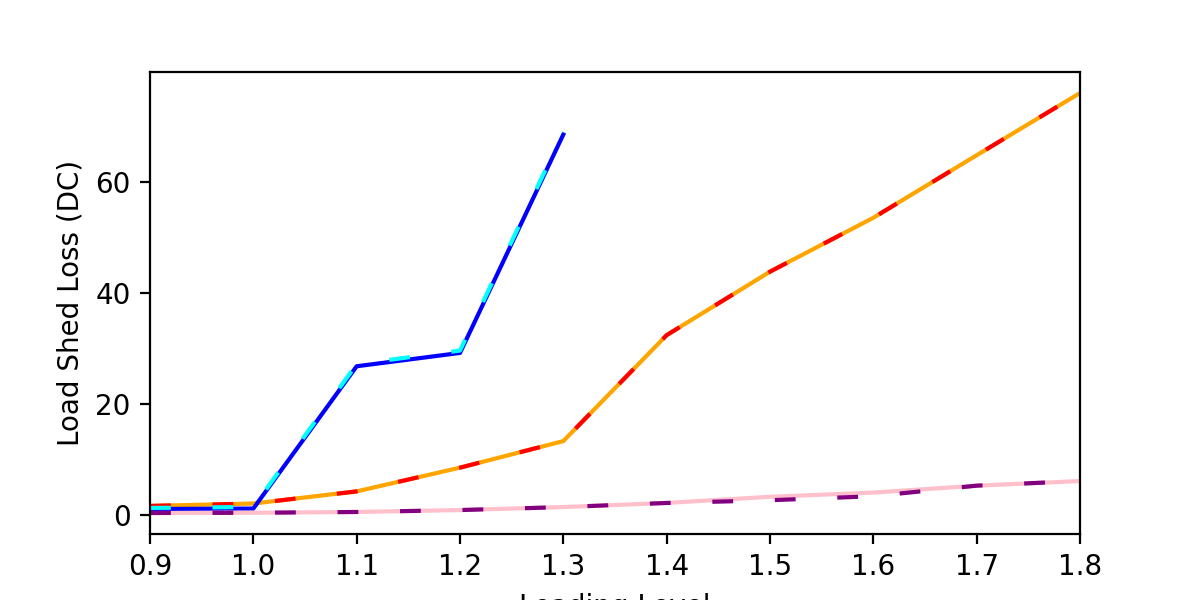}
\captionsetup{font={footnotesize}}\caption{Load Shed Loss (DC).} 
\label{fig: loadshed DC}
\end{minipage}\hfill
\begin{minipage}{0.22\textwidth}
\includegraphics[width=1\textwidth]{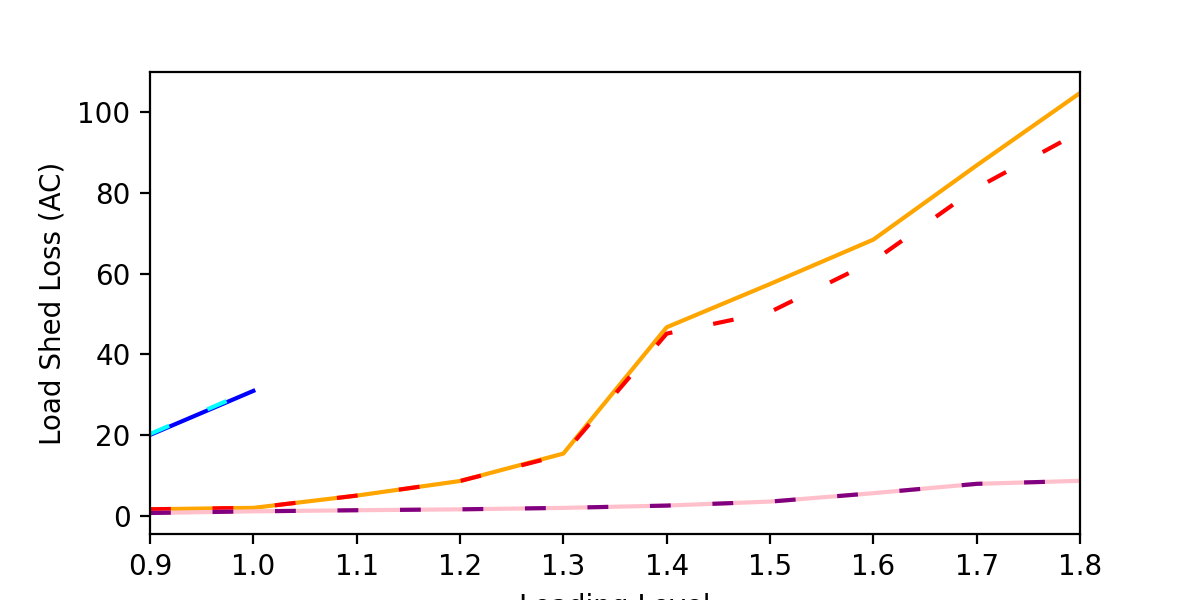}
\captionsetup{font={footnotesize}}\caption{Load Shed Loss (AC).} 
\label{fig: loadshed AC}
\end{minipage}
\begin{minipage}{0.22\textwidth}
\includegraphics[width=1\textwidth]{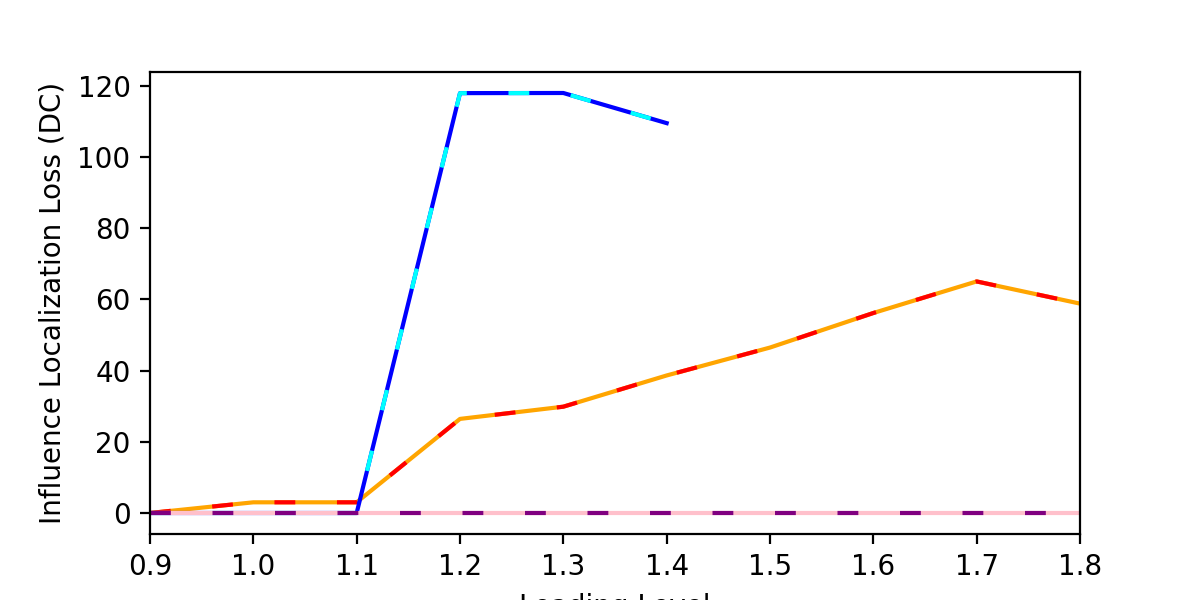}
\captionsetup{font={footnotesize}}\caption{Influence Localization Loss (DC).} 
\label{fig: influence loss DC}
\end{minipage}\hfill
\begin{minipage}{0.22\textwidth}
\includegraphics[width=1\textwidth]{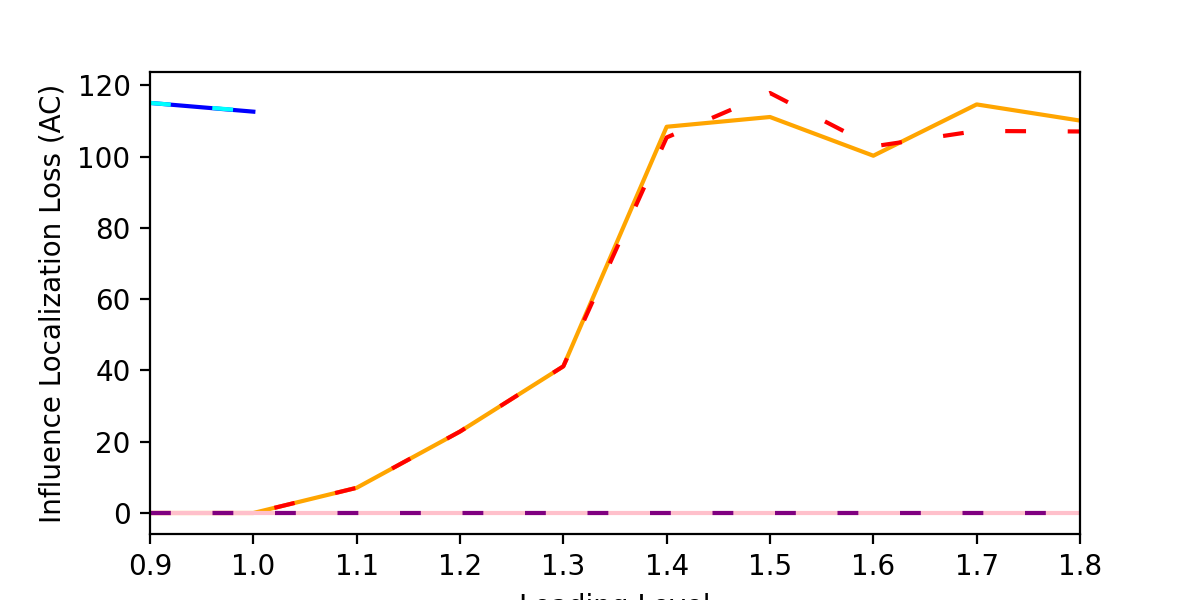}
\captionsetup{font={footnotesize}}\caption{Influence Localization Loss (AC).}
\label{fig: influence loss AC}
\end{minipage}
\centering
\begin{minipage}{0.3\textwidth}
\includegraphics[width=1\textwidth]{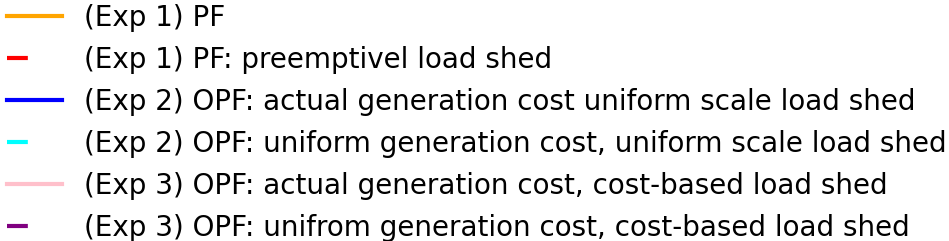}
\end{minipage}
\captionsetup{font={footnotesize}}\caption{Grid-centric, consumer-centric, and influence localization losses over under various corrective actions for DC and AC models.}
\label{fig:all plots}
\end{figure}

In \textit{Experiment 3} influence is entirely localized. In \textit{Experiment 1 \& 2}, influence is completely localized at low loading levels ($0.9\times$), and becomes increasingly less so as loading increase. Influence is significantly less contained for loading levels $>1.2\times,$ as seen in the spike in \cref{fig: influence loss DC} before the curve plateaus. Unlike losses on link failure or load shed, high loading levels does not imply greater influence localization loss. Similarly, we observed no significant difference between re-dispatching with actual or uniform generation cost.

\section{EFFECTS OF CORRECTIVE ACTIONS (AC MODELS)}\label{section 6: corrective actions AC}

We similarly analyze results with AC models. All trends in DC holds with the minor exceptions. Specifically, AC PF finds no cascade failure at the nominal loading level, as opposed to sparse cascades under DC PF. As seen in \cref{fig: influence loss DC}, the sharp transition in influence localization loss between $1.1\times $ and $1.2\times$ loading level in \textit{Experiment 3} remains.

\begin{figure}
\centering
\includegraphics[width=0.4\textwidth]{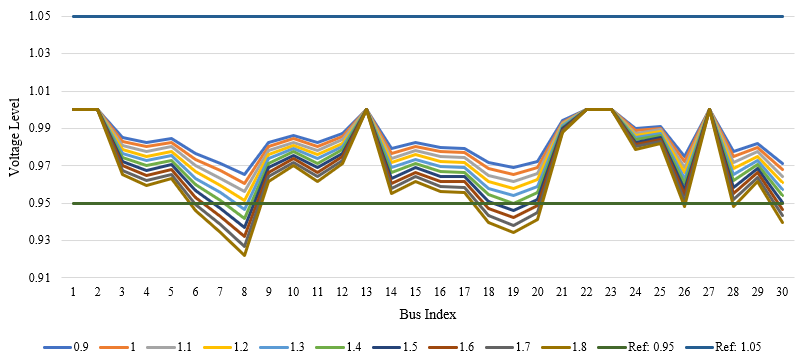}
\captionsetup{font={footnotesize}}\caption{Initial bus voltages solved with AC power flow.} \label{fig: ACPF}
\end{figure}

We find that PF solutions are frequently not physically implementable. This can be observed in the initial voltages under AC PF in \cref{fig: ACPF}. Bus voltages drops significantly as loading increases, falling outside the $(0.95, 1.05)$ constraint.

In all three experiments, AC models uncover more link failures, greater load shed, and lower levels of influence localization than their the DC model counterparts. In particular, in \textit{Experiment 1}, losses on link failure is only slightly greater in AC than DC models, but this difference is much greater in \textit{Experiment 2}. Losses on influence localization is much worse than the DC approximation in most cases. The high levels of losses from AC solutions renders the DC approximation insufficient to study failure cascades, as it underestimates the severity of contingencies.

Findings about the load shed losses present an especially optimistic outlook. In \textit{Experiment 1 \& 2}, AC simulations render much higher load shed than DC simulations. However, in \textit{Experiment 3}, when cost-based flexible load shed is implemented, this load shed cost is not longer so significant. As AC simulation results are physically implementable, this result shows that we can serve close to full demand without causing congestion with optimal re-dispatch for both generation and load. This is especially promising in practice.

\section{PERFORMANCE EVALUATIONS} \label{section 7: performance eval}

In this section, we outline the algorithm used for link failure and load shed prediction, propose evaluations metrics, and present our algorithm's performance. In each experiment, we generate 300 samples, of which 270 ($90\%$) are used for training and 30 ($10\%$) used for testing, repeating for each loading level ($0.9\times, 1.0\times, \dots, 1.8\times$ \cite{MATPOWER}'s default).

\subsection{Algorithm for Prediction} \label{section 7.1: algorithm for prediction}

We make predictions with bisection thresholds $\epsilon^{k^*}, \delta^{k^*}$ selected from Step 3 of \cref{section 2.4 epsilon}. We compare results from \cref{eq:s[t+1]} and compare them with $\epsilon^{k^*}.$ Links with probabilities greater than their threshold values are predicted to be alive. We make the improvement from \cite{Wu} by using the binary prediction of each step as inputs to \cref{eq:s[t+1]}, rather than the probabilities. This can be seen as applying an excitation function at each state of the Markov process, which significantly improved prediction accuracy.

To predict load shed, we use \cref{eq: l[t]} and similarly, compare the results with $\delta^{k^*}.$ Note that we use the true link status in our experiments, so that performance of the $D$ and $E$ influence models stay uncorrelated. This is impossible for real-life predictions, where true link status is unattainable.

\subsection{Evaluation Metrics} \label{section 7.2: eval metrics}

To evaluate link failure prediction accuracy, we adopt the link-based evaluations metric from \cite{Wu} and calculate the average final state prediction accuracy rate among all links. To evaluate load shed prediction accuracy, we establish a bus-wise metric to capture the prediction on any specific bus. Bus-specific inquiries can help identify areas at higher blackout risk. The accuracy score for test case $k$ is determined by \begin{equation} \label{eq: load shed metric}
    accuracy ^k = ||overall^k - \hat{overall}^k||_1,
\end{equation}
where $overall^k$ is a vector of length $N_{br}$ where the $i$-th entry is 1 if a load shed ever occurred at bus $i$ during the real cascade, and $0$ otherwise. The vector $\widehat{overall}^k$ holds similar information for the predicted load shed sequence $\widetilde{l}^k.$ We obtain a mean prediction accuracy over all test samples.

\begin{figure}
\begin{minipage}{0.24\textwidth}
\includegraphics[width=1\textwidth]{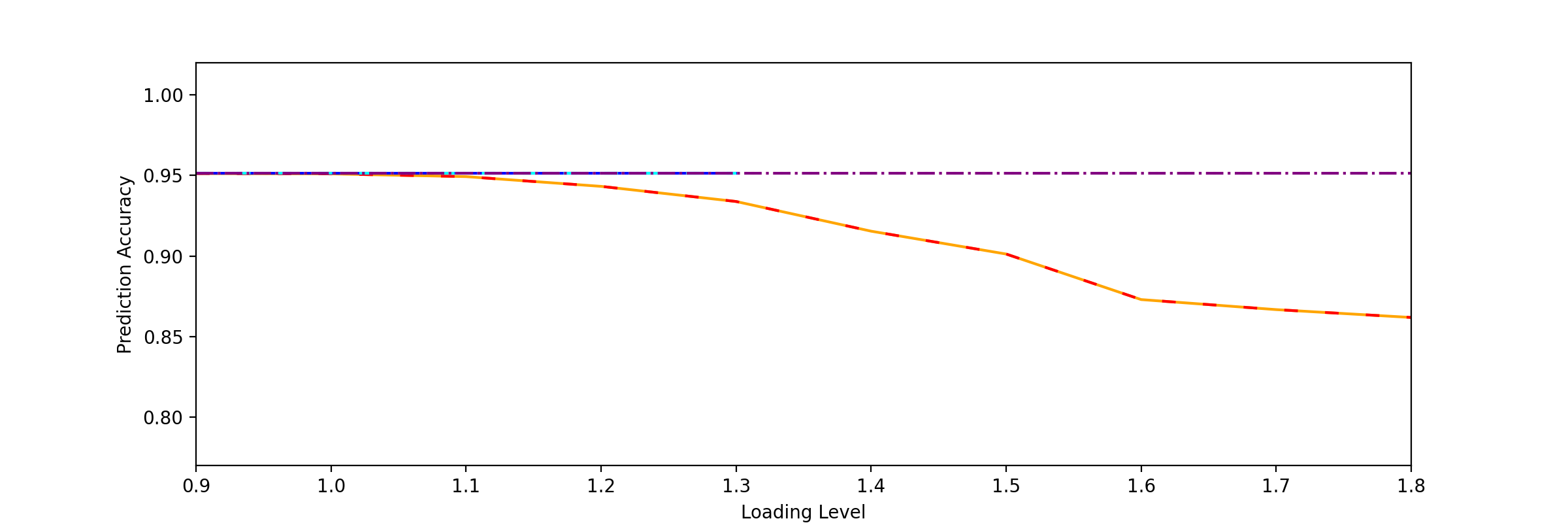}
\captionsetup{font={footnotesize}}\caption{Link Failure (DC training).} 
\label{fig: link fail train DC}
\end{minipage}\hfill
\begin{minipage}{0.24\textwidth}
\includegraphics[width=1\textwidth]{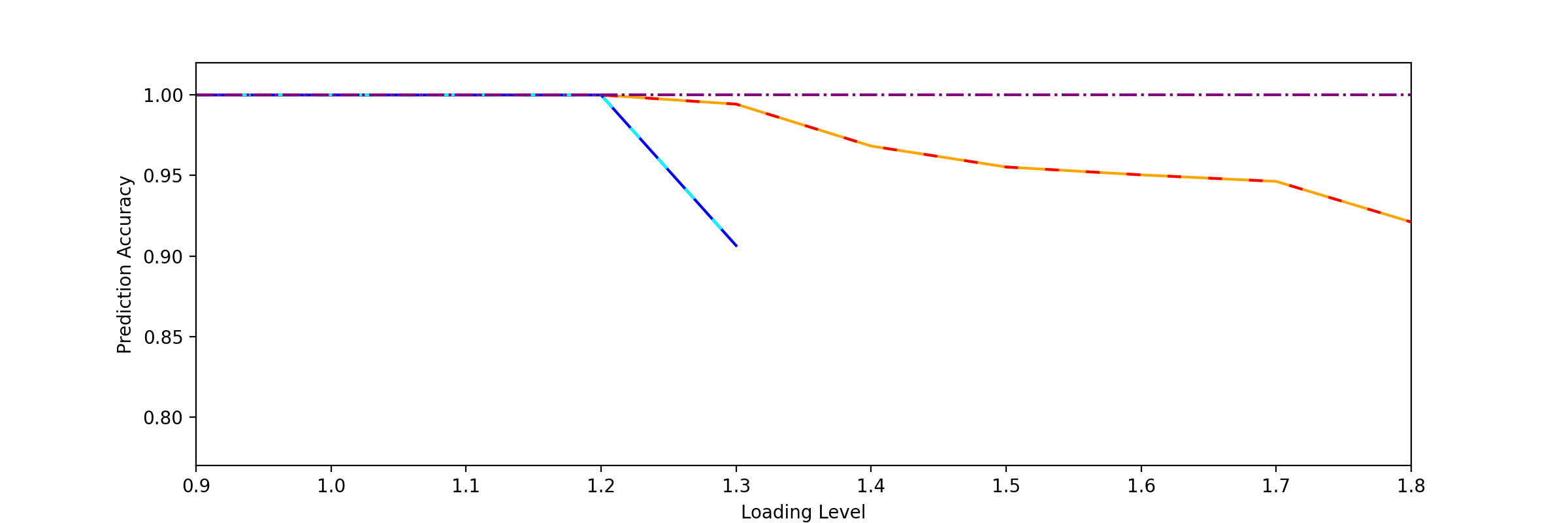}
\captionsetup{font={footnotesize}}\caption{Link Failure (DC testing).} 
\label{fig: link fail test DC}
\end{minipage}
\begin{minipage}{0.24\textwidth}
\includegraphics[width=1\textwidth]{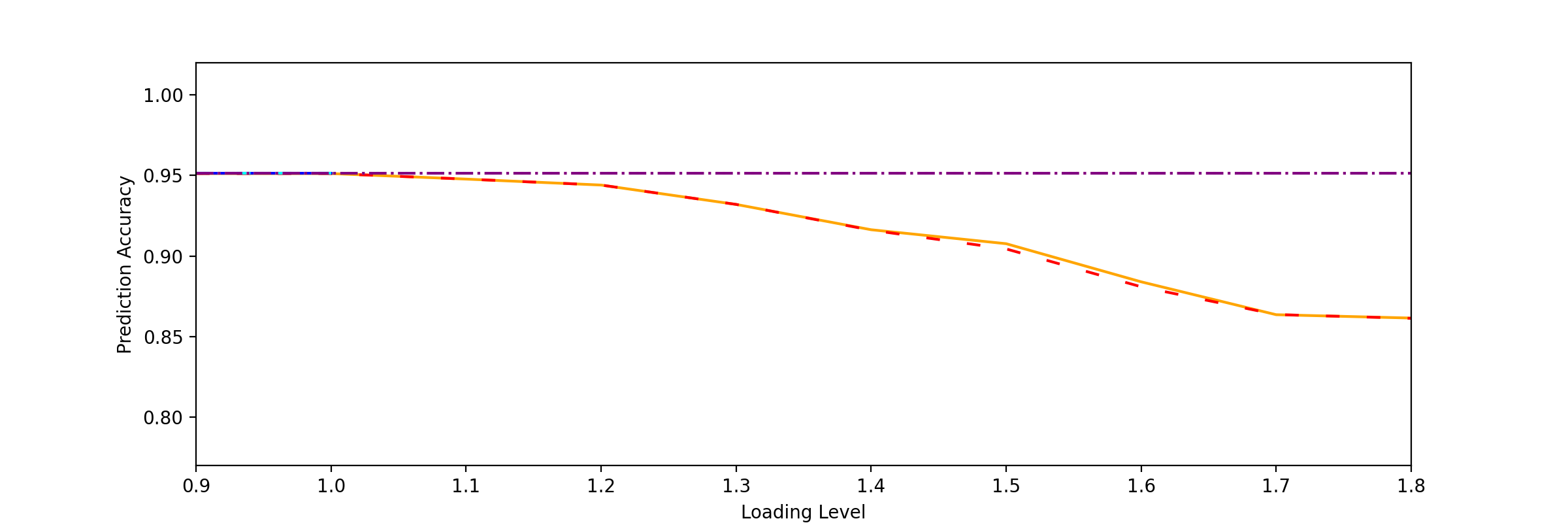}
\captionsetup{font={footnotesize}}\caption{Link Failure (AC training).} 
\label{fig: link fail train AC}
\end{minipage}\hfill
\begin{minipage}{0.24\textwidth}
\includegraphics[width=1\textwidth]{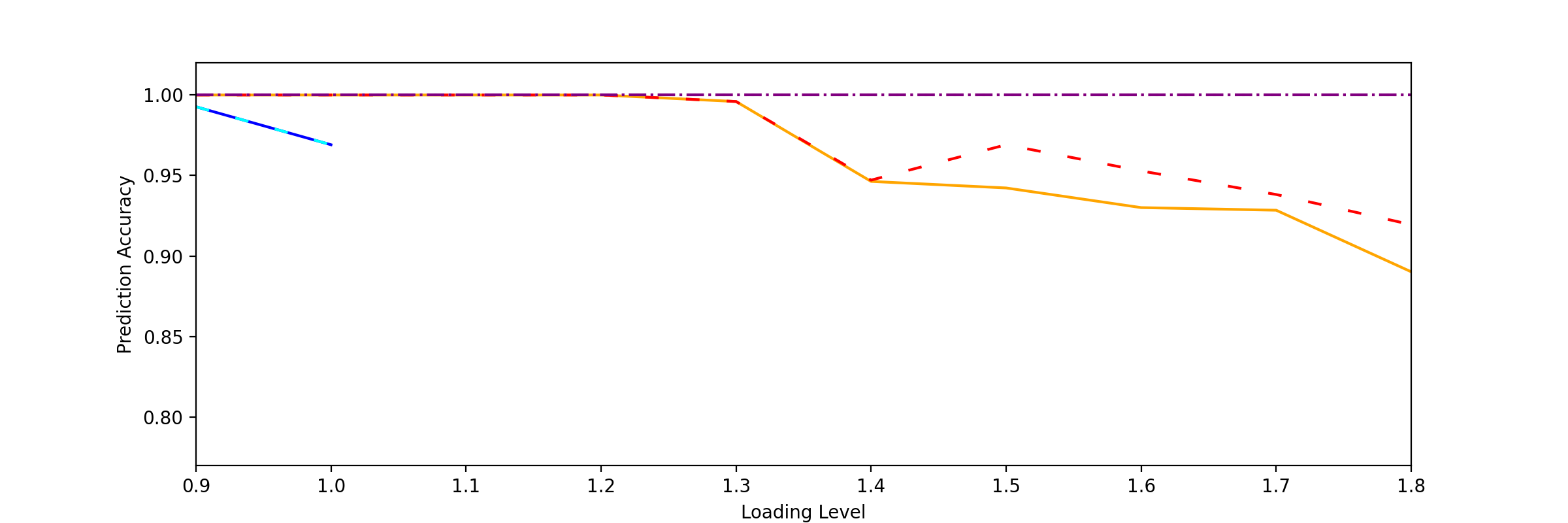}
\captionsetup{font={footnotesize}}\caption{Link Failure (AC testing).} 
\label{fig: link fail test AC}
\end{minipage}
\begin{minipage}{0.24\textwidth}
\includegraphics[width=1\textwidth]{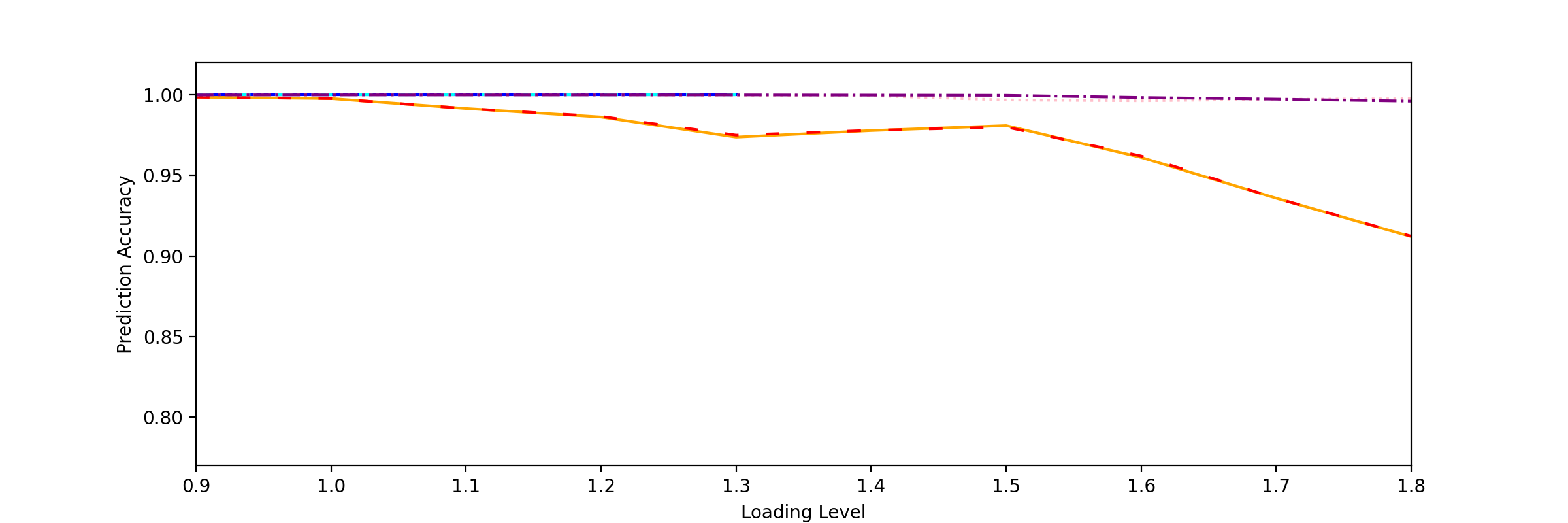}
\captionsetup{font={footnotesize}}\caption{Load Shed (DC training).} 
\label{fig: load shed train DC}
\end{minipage}\hfill
\begin{minipage}{0.24\textwidth}
\includegraphics[width=1\textwidth]{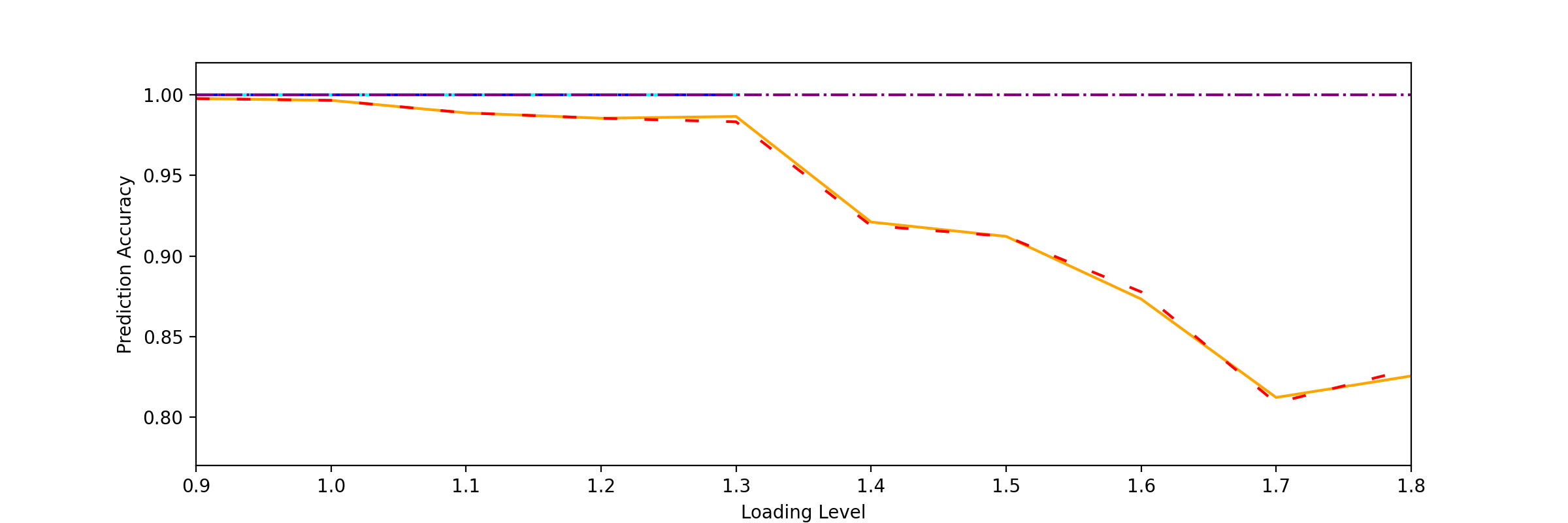}
\captionsetup{font={footnotesize}}\caption{Load Shed (DC testing).} 
\label{fig: load shed test DC}
\end{minipage}
\begin{minipage}{0.24\textwidth}
\includegraphics[width=1\textwidth]{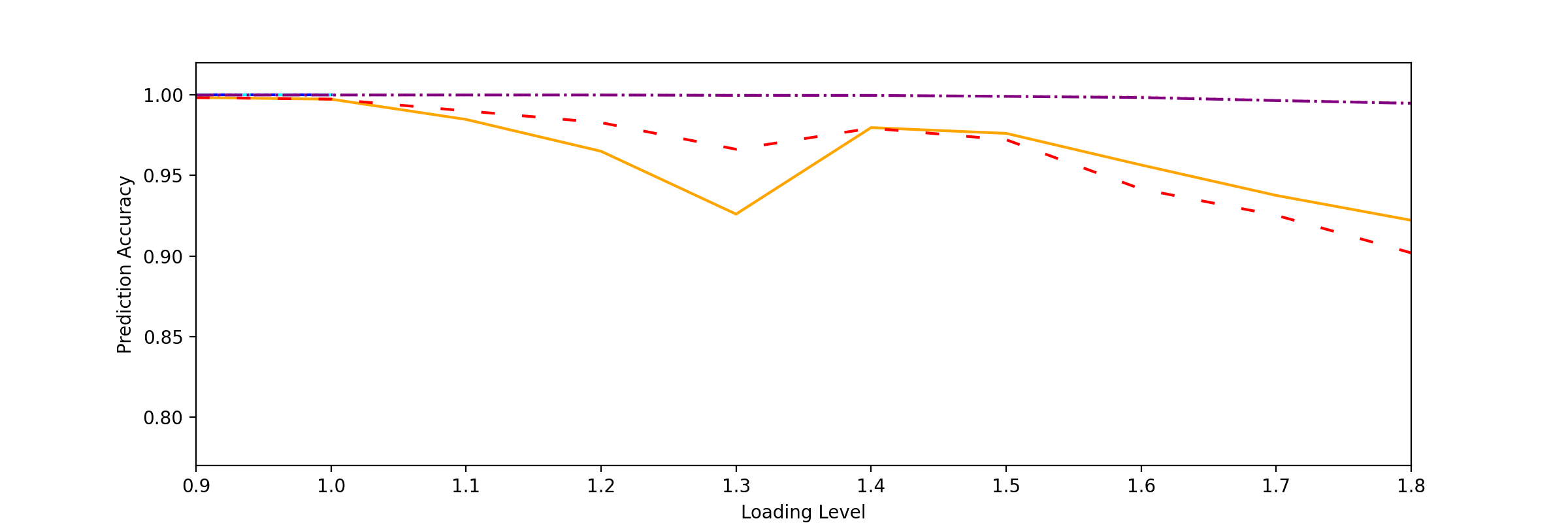}
\captionsetup{font={footnotesize}}\caption{Load Shed (AC training).} 
\label{fig: load shed train AC}
\end{minipage}\hfill
\begin{minipage}{0.24\textwidth}
\includegraphics[width=1\textwidth]{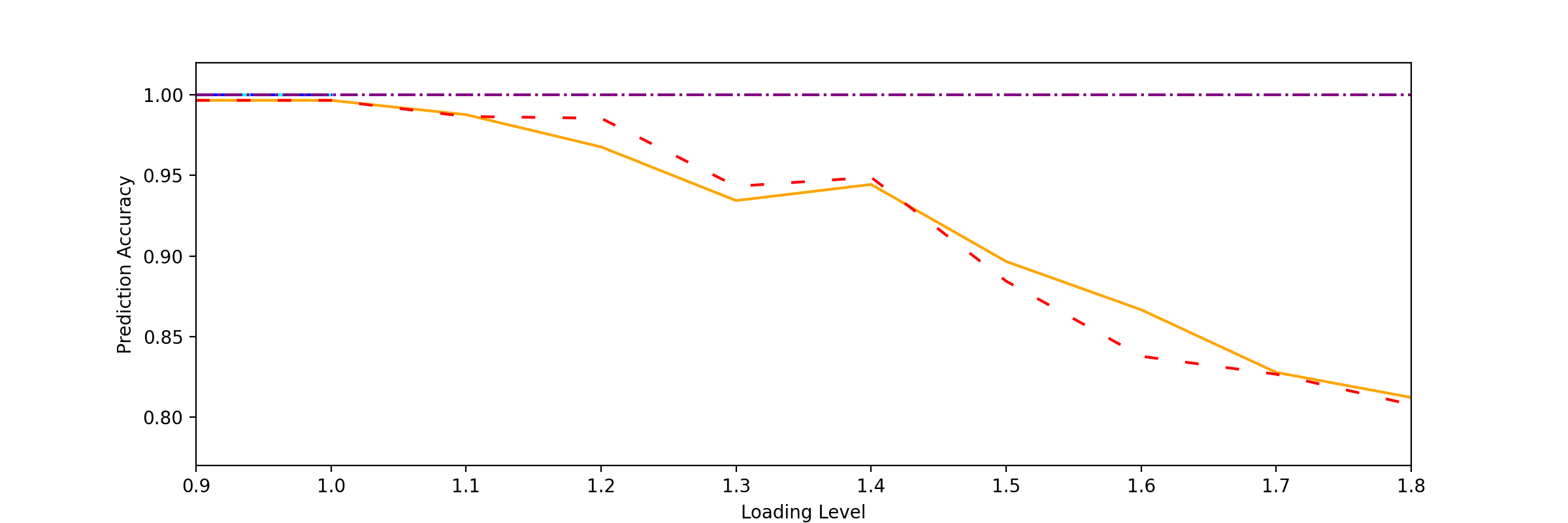}
\captionsetup{font={footnotesize}}\caption{Load Shed (AC testing).} 
\label{fig: load shed test AC}
\end{minipage}
\centering
\begin{minipage}{0.3\textwidth}
\includegraphics[width=1\textwidth]{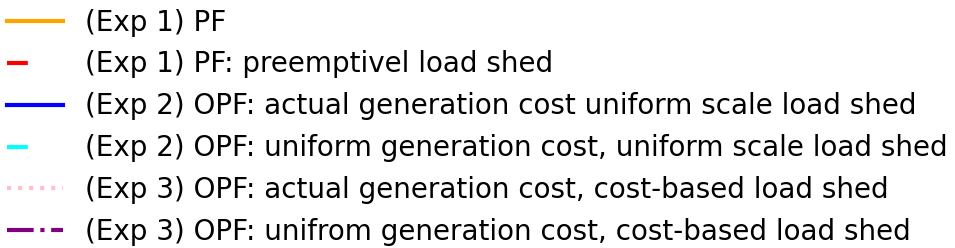}
\end{minipage}
\captionsetup{font={footnotesize}}\caption{Link Failure and Load Shed Prediction Accuracy.}
\label{fig:all predictions}
\end{figure}

\subsection{Performance Analysis} \label{section 7.3: performance analysis}

\cref{fig: link fail train DC} - \cref{fig: link fail test AC} present the results on link-based failure prediction accuracy on the training and testing sets  over loading levels varying from $0.9\times$ to $1.8\times.$ \cref{fig: load shed train DC} - \cref{fig: load shed test AC} present bus-based accuracy for load shed prediction.

We highlight two observations. First, both tasks reach high prediction accuracy, $>90\%$ for most cases and $>80\%$ for all. There is no significant difference between the training and testing sets for link failure predictions, which verifies that our model does not overfit the training sample. As for load shed predictions, accuracy over the training set is higher than testing set. We find no significant difference on prediction performance between the DC and AC models.

Next, we demonstrate the superiority of the influence model framework to the power flow-based contingency calculation by CFS in computational time cost reduction under both DC and AC flow models. Our findings corroborate \cite{Wu} Consider a sample of life time $\tau.$ To deterministically find the link overflows and load sheds, we need to run PF or OPF once for each disconnected component at each time step, which totals to at least $\tau$ times and likely much more. On the other hand, under our influence model, each time step consists of one matrix multiplication and one comparison step for both link failure and load shed prediction, significantly reducing computational cost. This is especially significant for the AC models, as it avoids solving nonlinear equations. Additionally, computational efficiency is identical for all cases under our method, but flow-based solutions become more computationally expensive as loading level increases to causes more failures and disconnections in the grid.

While our experiments are ran over the IEEE-30 system, we believe that the cost reduction will be more significant for larger systems, as non-linear optimization because exponentially expensive with more decision variables and constraints.

\subsection{Structures of System-Wide Influence} \label{section 7.4: structures}

\begin{figure}
\begin{minipage}{0.15\textwidth}
\includegraphics[width=1\textwidth]{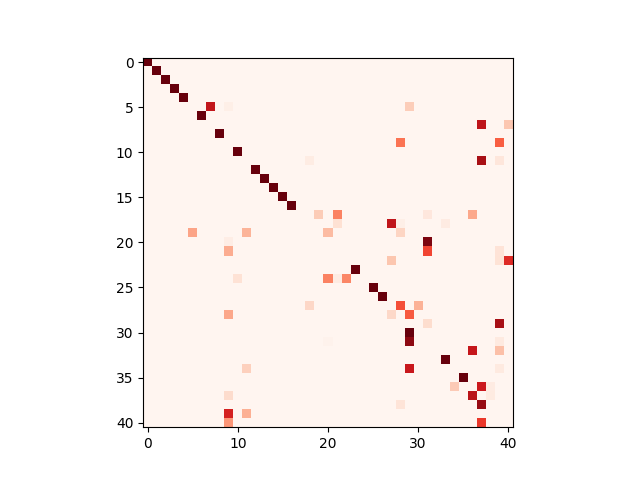}
\captionsetup{font={footnotesize}}\caption{$D$ matrix for DC PF, $1.6\times$ loading} 
\label{fig: D exp1-DC}
\end{minipage}\hfill
\begin{minipage}{0.15\textwidth}
\includegraphics[width=1\textwidth]{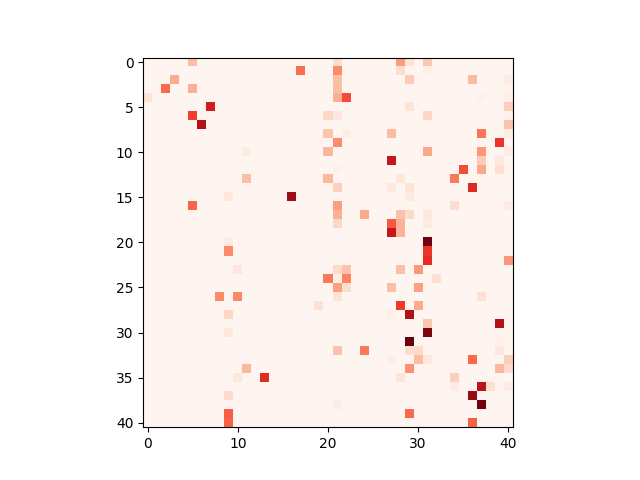}
\captionsetup{font={footnotesize}}\caption{$D$ matrix for AC PF, $1.6\times$ loading} 
\label{fig: D exp1-AC}
\end{minipage} \hfill
\begin{minipage}{0.15\textwidth}
\includegraphics[width=1\textwidth]{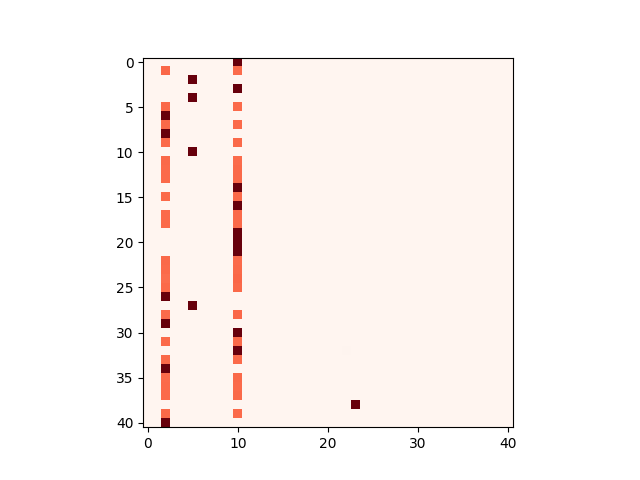}
\captionsetup{font={footnotesize}}\caption{$D$ matrix for DCOPF, $1\times$ loading} 
\label{fig: D exp2-DC}
\end{minipage} \\
\begin{minipage}{0.15\textwidth}
\includegraphics[width=1\textwidth]{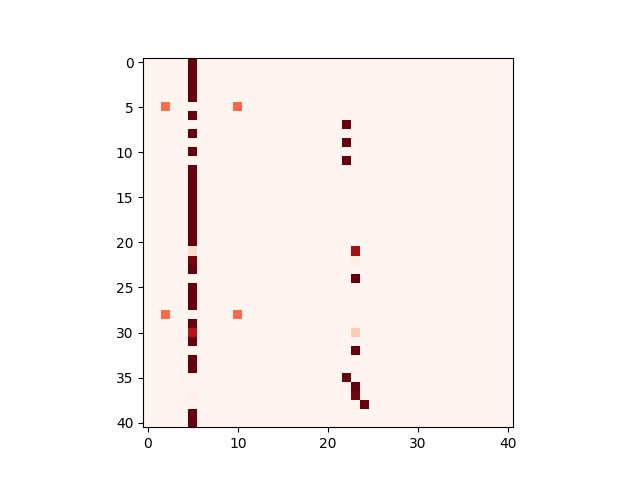}
\captionsetup{font={footnotesize}}\caption{$D$ matrix for ACOPF, $1\times$ loading} 
\label{fig: D exp2-AC}
\end{minipage} \hfill
\begin{minipage}{0.15\textwidth}
\includegraphics[width=1\textwidth]{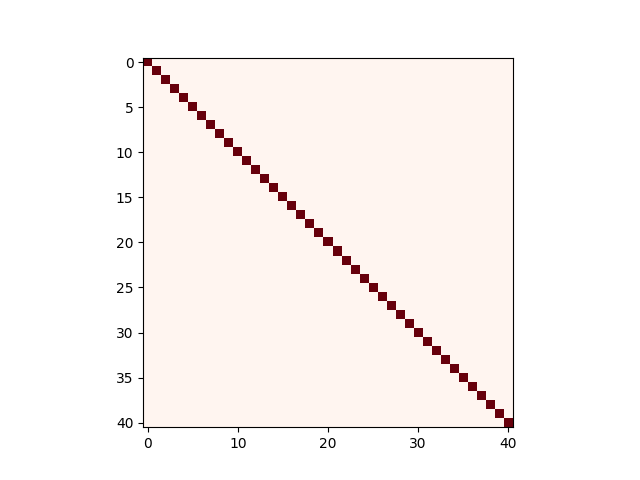}
\captionsetup{font={footnotesize}}\caption{$D$ matrix for DCOPF with smart load shed, $1.3\times$ loading} 
\label{fig: D exp3-DC}
\end{minipage}\hfill
\begin{minipage}{0.15\textwidth}
\includegraphics[width=1\textwidth]{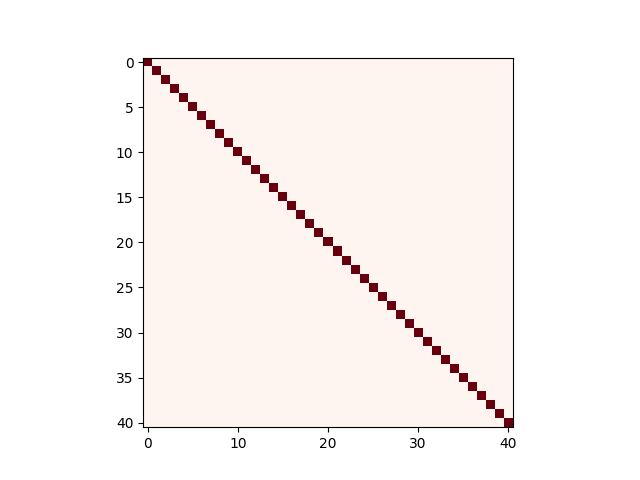}
\captionsetup{font={footnotesize}}\caption{$D$ matrix for ACOPF with smart load shed, $1.3\times$ loading} 
\label{fig: D exp3-AC}
\end{minipage}
\captionsetup{font={footnotesize}}\caption{$D$ matrix structures.}
\label{fig:D}
\end{figure}

\subsubsection{$D$ Matricies} \label{section 7.4.1: D structures}
A few interesting system-wide influence structures arise from our $D$ matrices. \cref{fig:D} shows a heat map plotted on the influence matrix $D$ for selected scenarios, where darker colors denote higher influence levels. \cref{fig: D exp1-DC} is an example prototypical for DC models when no corrective actions are taken, we observe to have a sparse structure. We may exploit this structure to reduce the time costs in \cref{section 2.3 estimating D} to obtain $D,$ which signals that our methodology is practical on larger, more complex networks.

However, when corrective actions are taken, the $D$ matrices display a linear structure, where the pair-wise influence values are high in particular columns (\cref{fig: D exp2-DC},\cref{fig: D exp2-AC}). This suggests a uni-directional, strong influence from one to many other links. $D$ matrices from \textit{Experiment 3} are entirely diagonal(\cref{fig: D exp3-DC}, \cref{fig: D exp3-AC}), as no link has failed.

Influence is found to be linear for AC models. \cref{fig: D exp1-AC} plots a prototypical matrix $D$ heat map for AC models. This trend is consistent for \textit{Experiment 1 \& 2}. \textit{Experiment 3}'s influence pattern is entirely diagonal (\cref{fig: D exp3-AC}) as no failure propagates, consistent with the DC case.

The linear structure found in our models has significant practical value. We may identify critical links that correspond to columns in $D$ with high influences. This can be extremely informative to system operators: when critical links fail, there is higher value to execute optimal generation re-dispatch and load shed according to the scheme in \textit{Experiment 3} so as to preserve infrastructure integrity and prevent even more failures and blackouts. 

\begin{figure}
\begin{minipage}{0.15\textwidth}
\includegraphics[width=1\textwidth]{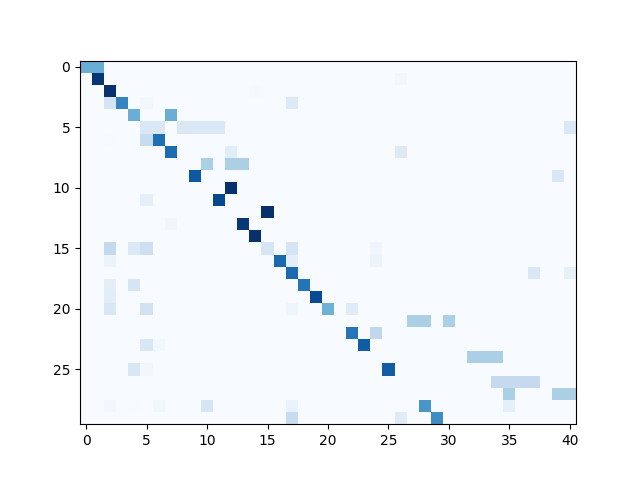}
\captionsetup{font={footnotesize}}\caption{$E$ matrix for DC PF, $1.6\times$ loading} 
\label{fig: E exp1-DC}
\end{minipage}\hfill
\begin{minipage}{0.15\textwidth}
\includegraphics[width=1\textwidth]{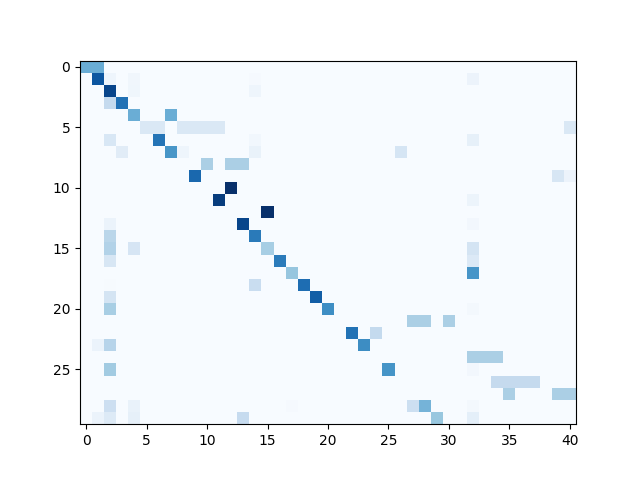}
\captionsetup{font={footnotesize}}\caption{$E$ matrix for AC PF, $1.6\times$ loading} 
\label{fig: E exp1-AC}
\end{minipage} \hfill
\begin{minipage}{0.15\textwidth}
\includegraphics[width=1\textwidth]{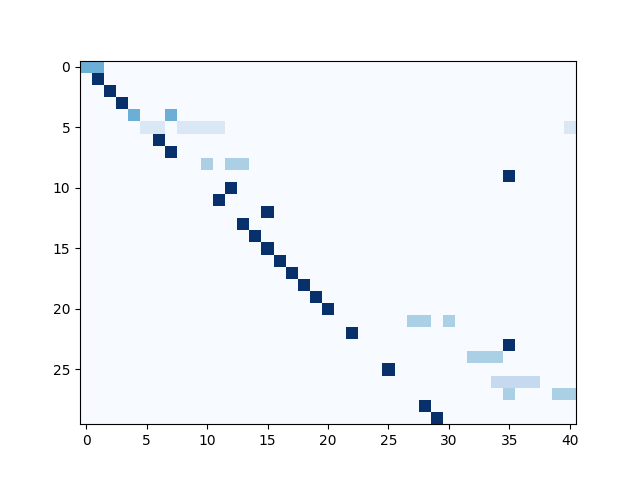}
\captionsetup{font={footnotesize}}\caption{$E$ matrix for DCOPF, $1\times$ loading} 
\label{fig: E exp2-DC}
\end{minipage}\hfill
\begin{minipage}{0.15\textwidth}
\includegraphics[width=1\textwidth]{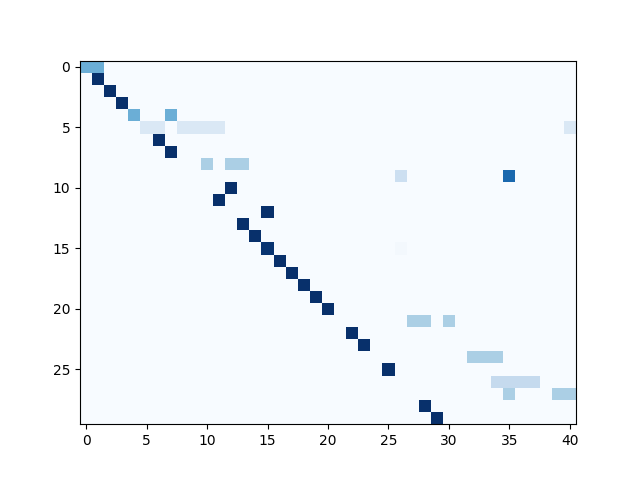}
\captionsetup{font={footnotesize}}\caption{$E$ matrix for ACOPF, $1\times$ loading} 
\label{fig: E exp2-AC}
\end{minipage} \hfill
\begin{minipage}{0.15\textwidth}
\includegraphics[width=1\textwidth]{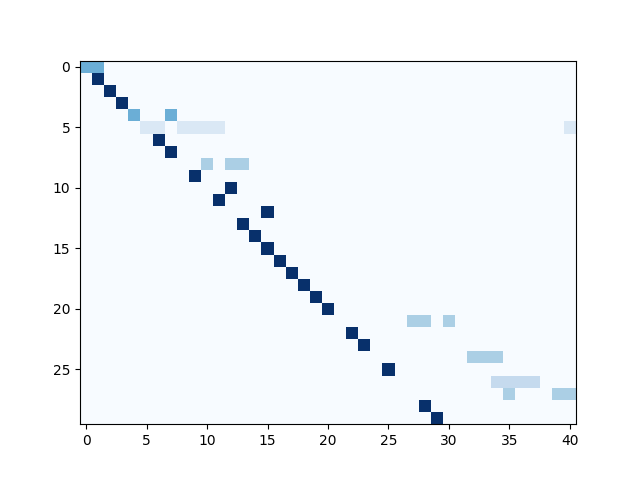}
\captionsetup{font={footnotesize}}\caption{$E$ matrix for DCOPF with smart load shed, $1.3\times$ loading} 
\label{fig: E exp3-DC}
\end{minipage}\hfill
\begin{minipage}{0.15\textwidth}
\includegraphics[width=1\textwidth]{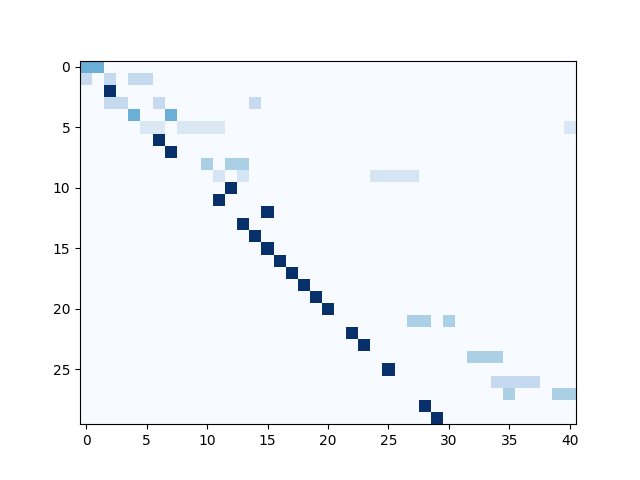}
\captionsetup{font={footnotesize}}\caption{$E$ matrix for ACOPF with smart load shed, $1.3\times$ loading} 
\label{fig: E exp3-AC}
\end{minipage}
\captionsetup{font={footnotesize}}\caption{$E$ matrix structures.}
\label{fig:E}
\end{figure}

\subsubsection{$E$ Matrices} \label{section 7.4.2: E structures}

Careful examination of the $E$ matrices give rise to a few other interesting findings. First, like the $D$ matrix, the structure of the $E$ matrices tend to be sparse (\cref{fig:E}), which allude to promises of lower time cost and scalability of learning the their values. It is even more so under \textit{Experiment 2 \& 3}, as illustrated in \cref{fig: E exp2-AC} and \cref{fig: E exp3-AC}, when generation re-dispatch is executed during the cascade. The higher levels of sparsity in \textit{Experiment 2} and \textit{Experiment 3} can be explained by their designs: generation re-dispatch as corrective actions reduces chance of failure in subsequent time steps, incurring load sheds in the cascade sequence. 

However, we also observed that different experiment setups produce similar $E$ matrices, with higher levels of sparsity with re-dispatching. Yet, many influences preserve. This signifies a deterministic relationship between certain failures and load shed at particular buses. While the deterministic relationship is pessimistic, it can very well be informative to system operators of potential service reductions.

\section{ADVISORY TOOL FOR OPERATORS} \label{section 8: advisory tool}

In this section, we propose a few practical applications to our influence model in risk evaluation, prediction, and corrective action advisory.

\subsection{Risk Evaluation} \label{section 8.1: risk evaluation}
The influence models helps us determine the most criticality links and the most risky initial contingencies.

We evaluate the criticality of each link by tallying the effect its failure has on all other links or buses. Grid and consumer-centric criticality values are computed with equations as follows.
\begin{equation}
    C^D(j) = \sum_{i=1}^{N_{br}} d_{ij}(A^{11}_{ji}-A^{01}_{ji}),
\end{equation} and
\begin{equation}
    C^E(j) = \sum_{i=1}^{N} e_{ij}(B^{11}_{ji}-B^{01}_{ji}),
\end{equation}
where $j$ is the link index, and $i$ enumerates over all links for $C^D(j)$ or all buses for $C^E(j)$.

With this information, system operators can identify the risks upon contingencies. Criticality values also lends insight into infrastructure maintenance --- links of higher criticality require more frequent checkup and maintenance.

Additionally, the link failure, load shed, and influence localization losses computed based on \cref{eq: link fail loss}, \cref{eq: load shed loss}, \cref{eq: local influence loss} may be used to inform system operators which combination of initial failures pose high risk to the network.

\subsection{Prediction and Corrective Action Advisory}\label{section 8.2: prediction and advisory}

Give snapshot of the current network, can predict failure cascade and load sheds. Following \cref{section 7.1: algorithm for prediction}'s procedure, our model can give failure cascade and load shed prediction of the current network profile. This prediction can be done for no corrective action and various defense strategies, and comparing their results obtained can inform operators the best course of action for loss minimization.

\section{CONCLUSIONS} \label{section 9: conclusion}
This study's contribution can be summarized as follows. 
\begin{enumerate}[leftmargin=*]
    \item We examine the effects of corrective actions during a failure cascade. 
    We propose three loss functions (grid-centric, consumer-centric, and influence localization) to evaluate the criticality of each initial contingency and effectiveness of each action. We apply the existing IM framework to predict line failures. Analyses are done with models that ensure physically implementable solutions.
    \item We propose a novel IM framework for mandatory load shed prediction and verify its prediction accuracy and computation efficiency under different scenarios. 
    \item We propose an advisory tool to inform operators of failure criticality, cascade and load shed prediction, and corrective action recommendation.
\end{enumerate}

We illustrate our models on the IEEE-30 system. Promising results render us confidence in the robustness of our framework in even larger networks.

\section*{ACKNOWLEDGMENTS}

The first author appreciates the MIT UROP and MIT Energy Initiative for funding. The second author appreciates partial funding by the NSF EAGER project \#2002570. We also thank Dan Wu, Xinyu Wu, and Miroslav Kosanic for having meaningful discussions with us.


\begin{thebibliography}{99}

\bibitem{bg-1}
``NorthEast US Failure Cascade,'' https://www.bostonglobe.com/magazine/2012/02/03/anatomy-blackout-august/mAsrr41nLAjGFIU3IF440O/story.html.
\bibitem{bg-2}  ``Manhattan, New York Failure Cascade,'' https://www.theatlantic.com/technology/archive/2019/07/manhattan-blackout-reveals-infrastructure-risk/594025/.
\bibitem{bg-3} “London Failure Cascade,” https://www.bloomberg.com/news/articles/2019-08-09/london-blackout-occurred-amid-drop-in-wind-and-natural-gas-power.
\bibitem{yamashita} Yamashita, K., Joo, S.-K., Li, J., Zhang, P. and Liu, C.-C. (2008), Analysis, control, and economic impact assessment of major blackout events. Euro. Trans. Electr. Power, 18: 854-871. https://doi.org/10.1002/etep.304
\bibitem{puerto_rico} Ilic, M., Ulerio, R. S., Corbett, E., Austin, E., Shatz, M., \& Limpaecher, E. (2020). A Framework for Evaluating Electric Power Grid Improvements in Puerto Rico.
\bibitem{earlyworks} Dobson I., Carreras B., Lynch V., Newman D., "Complex systems analysis of series of blackouts: Cascading failure, critical points, and self-organization", Chaos 17, 026103 (2007) https://doi.org/10.1063/1.2737822
\bibitem{allen} S. Cvijić, M. Ilić, E. Allen and J. Lang, "Using Extended AC Optimal Power Flow for Effective Decision Making," 2018 IEEE PES Innovative Smart Grid Technologies Conference Europe (ISGT-Europe), 2018, pp. 1-6, doi: 10.1109/ISGTEurope.2018.8571792.
\bibitem{CFS} M. J. Eppstein and P. D. H. Hines, "A “Random Chemistry” Algorithm for Identifying Collections of Multiple Contingencies That Initiate Cascading Failure," in IEEE Transactions on Power Systems, vol. 27, no. 3, pp. 1698-1705, Aug. 2012, doi: 10.1109/TPWRS.2012.2183624.
\bibitem{vaiman} Vaiman et al., "Risk Assessment of Cascading Outages: Methodologies and Challenges," in IEEE Transactions on Power Systems, vol. 27, no. 2, pp. 631-641, May 2012, doi: 10.1109/TPWRS.2011.2177868.
\bibitem{history} M.B. Cain, R.P. O’neill and A. Castillo, "History of optimal power flow and formulations,'' 2012. Federal Energy Regulatory Commission, 1, pp. 1-36.
\bibitem{MATPOWER} R. D. Zimmerman, C. E. Murillo-Sánchez and R. J. Thomas, "MATPOWER: Steady-State Operations, Planning, and Analysis Tools for Power Systems Research and Education," in IEEE Transactions on Power Systems, vol. 26, no. 1, pp. 12-19, Feb. 2011, doi: 10.1109/TPWRS.2010.2051168.
\bibitem{sinha} M. Sinha, M. Panwar, R. Kadavil, T. Hussain, S. Suryanarayanan, and M. Papic. 2019. Optimal Load Shedding for Mitigation of Cascading Failures in Power Grids. In Proceedings of the Tenth ACM International Conference on Future Energy Systems (e-Energy '19). Association for Computing Machinery, New York, NY, USA, 416–418. https://doi.org/10.1145/3307772.3330172.
\bibitem{r-n} M. Rahnamay-Naeini, Z. Wang, N. Ghani, A. Mammoli and M. M. Hayat, "Stochastic Analysis of Cascading-Failure Dynamics in Power Grids," in IEEE Transactions on Power Systems, vol. 29, no. 4, pp. 1767-1779, July 2014, doi: 10.1109/TPWRS.2013.2297276.
\bibitem{ShiLiu} Benyun Shi, Jiming Liu, Decentralized control and fair load-shedding compensations to prevent cascading failures in a smart grid, International Journal of Electrical Power \& Energy Systems,
Volume 67, 2015, Pages 582-590, ISSN 0142-0615, https://doi.org/10.1016/j.ijepes.2014.12.041.
\bibitem{Cetinay} Hale Cetinay, Saleh Soltan, Fernando A. Kuipers, Gil Zussman, and Piet Van Mieghem. 2018. Analyzing Cascading Failures in Power Grids under the AC and DC Power Flow Models. SIGMETRICS Perform. Eval. Rev. 45, 3 (December 2017), 198–203. https://doi.org/10.1145/3199524.3199559.
\bibitem{Wu} X. Wu, D. Wu and E. Modiano, "Predicting Failure Cascades in Large Scale Power Systems via the Influence Model Framework," in IEEE Transactions on Power Systems, vol. 36, no. 5, pp. 4778-4790, Sept. 2021, doi: 10.1109/TPWRS.2021.3068409.
\bibitem{correction1}  Liu, Y., Wang, T., Gu, X.: A risk-based multi-step corrective control method for mitigation of cascading failures. IET Gener. Transm. Distrib. 16, 766– 775 (2022). https://doi.org/10.1049/gtd2.12327.
\bibitem{correction2} Shenhao Yang, Weirong Chen, Xuexia Zhang, Yu Jiang,
Blocking cascading failures with optimal corrective transmission switching considering available correction time,
International Journal of Electrical Power \& Energy Systems, Volume 141, 2022, 108248, ISSN 0142-0615, https://doi.org/10.1016/j.ijepes.2022.108248.
\bibitem{song} J. Song, E. Cotilla-Sanchez, G. Ghanavati and P. D. H. Hines, "Dynamic Modeling of Cascading Failure in Power Systems," in IEEE Transactions on Power Systems, vol. 31, no. 3, pp. 2085-2095, May 2016, doi: 10.1109/TPWRS.2015.2439237.
\bibitem{bg-zhang1} X. Zhang, C. Zhan and C. K. Tse, "Modeling the Dynamics of Cascading Failures in Power Systems," in IEEE Journal on Emerging and Selected Topics in Circuits and Systems, vol. 7, no. 2, pp. 192-204, June 2017, doi: 10.1109/JETCAS.2017.2671354.
\bibitem{bg-zhang2} Ding-Xue Zhang, Dan Zhao, Zhi-Hong Guan, Yonghong Wu, Ming Chi, Gui-Lin Zheng,
Probabilistic analysis of cascade failure dynamics in complex network, Physica A: Statistical Mechanics and its Applications, Volume 461, 2016, Pages 299-309, ISSN 0378-4371,  https://doi.org/10.1016/j.physa.2016.05.059.
\bibitem{one-line-diagram} Q. -S. Jia, M. Xie and F. F. Wu, "Ordinal optimization based security dispatching in deregulated power systems," Proceedings of the 48h IEEE Conference on Decision and Control (CDC) held jointly with 2009 28th Chinese Control Conference, 2009, pp. 6817-6822, doi: 10.1109/CDC.2009.5400740.

\end{thebibliography}
\end{document}